\documentclass[preprint,aps,prd,showpacs,superscriptaddress,nofootinbib]{revtex4}
\usepackage{mathbbold}
\usepackage{amsfonts}
\usepackage{mathrsfs}
\usepackage{graphicx}
\usepackage{amsmath}
\usepackage{amssymb}
\usepackage{bm}
\usepackage{bbm}

\def\OMIT#1{}

\newcommand{\nn}{\nonumber}

\newcommand{\beq}{\begin{equation}}
\newcommand{\eeq}{\end{equation}}
\newcommand{\bqa}{\begin{eqnarray}}
\newcommand{\eqa}{\end{eqnarray}}

\begin{document}

\title{\mbox{}\\[10pt]
Refactorizing NRQCD short-distance coefficients in exclusive
quarkonium production}

\author{Yu Jia\footnote{E-mail: jiay@ihep.ac.cn}}
\affiliation{Institute of High Energy Physics, Chinese Academy of
Sciences, Beijing 100049, China\vspace{0.2cm}}
\affiliation{Theoretical Physics Center for Science Facilities,
Chinese Academy of Sciences, Beijing 100049, China\vspace{0.2cm}}

\author{Deshan Yang\footnote{E-mail: yangds@gucas.ac.cn}}
\affiliation{College of Physical Sciences, Graduate University of
Chinese Academy of Sciences, Beijing 100049, China\vspace{0.2cm}}

\date{\today}
\begin{abstract}

In a typical exclusive quarkonium production process, when the
center-of-mass energy, $\sqrt{s}$, is much greater than the heavy
quark mass $m$, large kinematic logarithms of $s/m^2$ will
unavoidably arise at each order of perturbative expansion in the
short-distance coefficients of the nonrelativistic QCD (NRQCD)
factorization formalism, which may potentially harm the perturbative
expansion. This symptom reflects that the hard regime in NRQCD
factorization is too coarse and should be further factorized. We
suggest that this regime can be further separated into ``hard" and
``collinear" degrees of freedom, so that the familiar light-cone
approach can be employed to reproduce the NRQCD matching
coefficients at the zeroth order of $m^2/s$ and order by order in
$\alpha_s$. Taking two simple processes, exclusive $\eta_b+\gamma$
production in $e^+ e^-$ annihilation and Higgs boson radiative decay
into $\Upsilon$, as examples, we illustrate how the leading
logarithms of $s/m^2$ in the NRQCD matching coefficients are
identified and summed to all orders in $\alpha_s$ with the aid of
Brodsky-Lepage evolution equation.
\end{abstract}

\pacs{\it 12.38.-t, 12.38.Cy, 12.39.St, 14.40.Gx}

\maketitle

\emph{Introduction.} One of the classical applications of
perturbative Quantum Choromodynamics (QCD) is the successful
description of many exclusive processes with large momentum transfer
using collinear factorization, that allows one to express the
scattering amplitude as the convolution of perturbatively calculable
short-distance parts and the nonperturbative but universal
light-cone distribution amplitudes
(LCDA)~\cite{Lepage:1980fj,Chernyak:1983ej}. In describing hard
exclusive processes involving heavy quarkonium, {\it i.e.}, a
nonrelativistic bound state made of a heavy quark and a heavy
antiquark, however, there also exists another widely-accepted
theoretical framework, the nonrelativistic QCD (NRQCD) factorization
formalism~\cite{Bodwin:1994jh}. In this approach, the amplitude can
also be put in a factorized form, that is, an infinite sum of
products of short-distance coefficients and nonperturbative, albeit
universal, NRQCD matrix elements.

In recent years, considerable amount of efforts have been spent to
understand exclusive charmonium production mechanisms from both
NRQCD and light-cone perspectives. This endeavor is perhaps largely
propelled by a somewhat unexpected finding, that the lowest-order
NRQCD calculation of double charmonium production rate for the
process $e^+ e^-\to J/\psi\eta_c$~\cite{Braaten:2002fi} fell short
by about one order-of-magnitude of the \textsc{Belle}
measurement~\cite{Abe:2002rb}. The validity of applying NRQCD to
charmonia was soon questioned by some authors, who advocated that if
charmonium is treated as light meson, the light-cone approach
instead could satisfactorily accommodate the \textsc{Belle}
data~\cite{Ma:2004qf,Bondar:2004sv,Braguta:2008tg}. However, a
careful reexamination~\cite{Bodwin:2006dm} suspected that these
optimistic assertions are premature and it was argued that a
``consistent" light-cone analysis would in fact yield a result not
much different from NRQCD, so that the situation has not truly
improved.

In reviewing this episode, one may get the impression that the
light-cone and NRQCD approaches are two drastically different, and,
competing, theoretical frameworks. Indeed, these two approaches are
rooted in two different types of operator-product-expansions (OPE),
where the former is intimately linked to the light-cone (twist)
expansion, and the latter is more closely related to a local (large
mass) expansion.

The main motif of this work is to show that, these two approaches,
both bearing solid theoretical grounds, need not to be regarded
solely as rivals. As a matter of fact, they can benefit each
other~\footnote{We note that there already exist some attempts along
this line, but motivated by somewhat different considerations. In
particular, there is a recent work trying to bridge the
leading-twist LCDA of $S$-wave charmonia with the local NRQCD matrix
elements~\cite{Ma:2006hc} (see also
\cite{Bell:2008er,Braguta:2006wr}).}. In particular, it turns out
that the concepts and techniques developed in the light-cone
approach can be fruitfully exploited to improve the NRQCD
calculations. For example, we will argue that, for some class of
exclusive single-quarkonium production processes, the NRQCD
short-distance coefficients can be reproduced order by order in
$\alpha_s$, in the language of collinear factorization. Moreover,
the light-cone approach can easily resum large kinematic logarithms
appearing in the NRQCD short-distance coefficients.

{\it Refactorization of NRQCD matching coefficients.} NRQCD is an
effective field theory that is constructed to portray the
nonrelativistic character of heavy quarkonium. The slow relative
motion between quark and antiquark in a quarkonium, $v$, naturally
constitutes the expansion parameter of the theory. To produce a
quarkonium, the heavy quark pair must be created within a distance
shorter than $1/m$ and have a small relative velocity, to warrant a
significant probability to form a quarkonium. The first condition
guarantees that the parton process can be calculated perturbatively
owing to asymptotic freedom. The latter requirement implies that the
full amplitude can be expanded in power of $v$, and NRQCD endows a
well-defined meaning for this Taylor expansion procedure.

For definiteness of our discussion, we will confine ourselves to the
{\it bottomonium} throughout this work, whenever a quarkonium is
referred to. Of course, all the results can be trivially copied to
the charmonium case. Furthermore, for simplicity, in this work we
will concentrate on the single quarkonium production process, and
not touch upon the double-quarkonium production for its additional
technical complication. For a typical hard process that creates a
bottomonium, generically denoted by $H$, NRQCD factorization demands
that the amplitude can be expressed as
%
\bqa
{\mathcal M}[H] &\sim&  \sum_n \mathcal{C}_n(Q,m) \langle H| {\cal
O}_n|0 \rangle,
\label{NRQCD:generic}
\eqa
%
where ${\cal O}_n$ stands for an appropriate local color-singlet
NRQCD operator, $m$ is the $b$ quark mass, $Q$ refers to a typical
kinematic scale such as the center-of-mass energy, presumably much
greater than $m$. The important feature is that each ${\cal O}_n$
has a definite power counting in $v$, so that the expansion in
(\ref{NRQCD:generic}) becomes practically maneuverable.

NRQCD factorization manifestly separates the effects of {\it hard}
degree of freedom ($p^\mu \gtrsim  m$) from the remaining
lower-energy ones~\footnote{There are totally three kinds of such,
{\it soft} ($p^\mu \sim mv$) , {\it potential} ($p^0\sim
mv^2,\:{\mathbf p}\sim mv$) and {\it ultrasoft} ($p^\mu \sim
mv^2$)~\cite{Brambilla:2004jw}. All these scalings are specified in
the quarkonium rest frame.}. In Eq.~(\ref{NRQCD:generic}), the
matching coefficients $\mathcal{C}_n$ encode the effects of hard
quantum fluctuation, which are perturbatively calculable and
infrared finite, whereas the local NRQCD matrix elements
characterize the long-distance, and primarily nonperturbative,
aspects of $H$.

The short-distance coefficient ${\mathcal C}_n$ depends on two
energy scales, $Q$ and $m$. When these two scales are widely
separated, it can be naturally organized by Taylor series in
$m^2/Q^2$, practically only the leading power term needs to be
retained~\footnote{In fact, for some classes of exclusive charmonium
production processes, it has been proved that NRQCD factorization
may no longer hold beyond the leading order in
$m^2/Q^2$~\cite{Bodwin:2008nf}. Therefore it seems not profitable,
for both pragmatic and theoretical reason, to proceed beyond the
leading power.}. It is always possible, by factoring out an
appropriate kinematic factor proportional to some powers of $Q$, to
render $\mathcal{C}_n$ dimensionless, and such that it only depends
on $m^2$ and $Q^2$ through their dimensionless ratio and it scales
with $m^2/Q^2$ at most logarithmically.

It is conceivable that at each order in loop expansion,
$\mathcal{C}_n$ receives contribution from logarithms of $m^2/Q^2$
as well as constants. When $Q\gg m$, there arises the ambiguity of
setting the renormalization scale in the $\alpha_s$, whether to put
it around $m^2$ or $Q^2$ may result in quite different answer for a
fixed order calculation. Moreover, these large logarithms may
potentially ruin the perturbative expansion in strong coupling
constant. Solution to these problems lies beyond the scope of NRQCD.
Needless to say, it is highly desirable that these large kinematic
logarithms in the NRQCD matching coefficients can be identified and
summed to all orders in $\alpha_s$, in order to ameliorate
perturbative expansion.

The cause for this symptom can be easily traced. The problem is that
the hard quanta integrated out by NRQCD still contains two widely
separated scales, accommodating the quantum fluctuations extending
from the scale $m$ to a much higher scale $Q$. It is natural to
speculate, for such a multi-scale problem, whether one can divide
this rather coarse region further into finer ones. To achieve this,
one needs first identify various relevant degrees of freedom, then
disentangle their contributions, and finally express $\mathcal{C}_n$
in a factorized form.

The NRQCD factorization is justified mainly because $b$ is heavy,
$m\gg \Lambda_{\rm QCD}$. As $Q \gg m$, the $H$ moves almost with
the speed of light in a natural reference frame such as the
center-of-mass frame. Since the {\it heavy} $b$ and $\bar{b}$ move
nearly along the light cone, they can be considered to be {\it
light} in the kinematic sense. We thus are facing a challenge where
the conflicting nature of ``light" heavy quarks must be consistently
incorporated.

From now on, we will specialize to the $S$-wave quarkonium
production. For simplicity, we will only consider the leading term
in the NRQCD expansion, since $\mathcal{O}_1$ is the simplest NRQCD
operator and constitutes the most important contribution. However,
we would like to stress that the purpose of imposing this
restriction is mainly for simplicity. Since our reasoning will be
based on quite general ground, there should be no principal
difficulty to proceed to higher orders in $v$ expansion.

To compute the short-distance coefficient $\mathcal{C}_1$, a
convenient method is to consider the quark amplitude where $H$ is
replaced by a free $b\bar{b}$ pair sharing the same quantum numbers
(color, spin, orbital angular momentum) as the leading Fock
component of $H$. At the lowest-order in $v$, $b$ and $\bar{b}$ are
forced to partition equally the total momentum of the pair, $P$. For
definiteness $P$ is supposed to move fast along the positive
$\hat{z}$ axis. Consider a generic loop diagram that contributes to
the corresponding quark amplitude. The so-called ``method of
region"~\cite{Beneke:1997zp} can be utilized to help identify the
relevant degrees of freedom in $\mathcal{C}_1$. There is a
``\emph{hard}" region~\footnote{We caution that the term \emph{hard}
here only refers to a subset of hard region in the sense of NRQCD,
and the actual meaning of this term may vary depending on different
context. Similarly the term ``soft" may also bear different meaning
in different places.}, in which the loop momentum scales as
$p^\mu\sim Q$. In this region, the $b$ quark can be treated as
massless. There are several kinds of infrared modes, \emph{soft}
($p^\mu\sim m$), \emph{collinear} ($p^+\sim Q, p^-\sim m^2/Q,
p_\perp\sim m$), and \emph{anti-collinear} ($p^+\sim m^2/Q, p^-\sim
Q, p_\perp\sim m$)~\footnote{Our convention in defining the
light-cone variable is $A^\pm={A^0\pm A^3\over \sqrt{2}}$, and a
four-vector $A^\mu$ is decomposed into $(A^+,A^-,{\mathbf A}_\perp)$
in this convention.}. All these infrared modes have the virtuality
of order $m^2$, hence the mass of $b$ quark must be retained in
these infrared regions. The validity of NRQCD factorization
guarantees there is no any overlap between these ``infrared" quanta
and those truly infrared modes intrinsic to NRQCD such as the
potential mode, as can be readily distinguished by their
virtualities.

It turns out that at the leading power of $m^2/Q^2$, the soft and
anti-collinear modes do not contribute to
$\mathcal{C}_1$~\cite{Jia:2009}. Therefore the relevant dynamic
degrees of freedom are only the hard and collinear modes, which
resembles the hard exclusive processes involving a single hadron
such as $\pi\!-\!\!\gamma$ transition form
factor~\cite{Lepage:1980fj}. Quite naturally, one can follow the
standard collinear factorization theorem at leading twist to cast
$\mathcal{C}_1$ as
\bqa
\mathcal{C}_1\left({m^2\over Q^2}\right) &\sim &  T(x,Q,\mu) \otimes
\hat{\phi}(x,m,\mu)+O(m^2/Q^2),
\label{lc-factorization}%
\eqa
where $T$ refers to the hard-scattering part involving massless
quarks, and $\hat{\phi}$ can be viewed as the LCDA of a free
$b\bar{b}$ pair with the same quantum numbers as $H$ (Later we will
give a rigorous definition for this LCDA). $x$ is the fraction of
plus-momentum carried by the $b$ quark with respect to that of the
pair, and the operation $\otimes$ implies the convolution over $x$
between two parts. $\mu$ is an arbitrary scale separating these two
regions. It is important to note that the scales $Q$ and $m$ now
become fully disentangled, {\it i.e.} the $T$ only depends on $Q$
but not on $m$, likewise the $\hat{\phi}$ cannot have any direct
sensitivity on the hard scale $Q$.

It is worth emphasizing that $\hat{\phi}$ introduced here (to be
concrete, at this moment let us specialize to the twist-2 LCDA of
the $b\bar{b}(^1S_0^{(1)},P)$ state), bears some important
difference with the twist-2 LCDA of a pion. The latter is a
genuinely nonperturbative object that is sensitive to the quantum
fluctuation at hadronic scale, so that it can only be extracted from
experiments or from some nonperturbative tools such as lattice
simulation or QCD sum rules. On the contrary, $m$ acts as the
infrared cutoff in the former case. Since $m$ is much greater than
$\Lambda_{QCD}$, $\hat{\phi}$ thus can be reliably computed in
perturbation theory. However, we note that, since both LCDAs of the
$b\bar{b}(^1S_0^{(1)},P)$ state and a pion possess the same
ultraviolet behavior, they obey the same renormalization group (RG)
equation.

The factorization formula (\ref{lc-factorization}) is expected to
hold to all orders in $\alpha_s$. Both of the two ingredients, $T$
and $\hat{\phi}$ are perturbatively improvable. When $Q\gg m$, this
light-cone-based method provides an alternative and efficient means
to reproduce the NRQCD short-distance coefficient, including the
logarithms as well as the constants, to any desired order in
$\alpha_s$.

At tree level, the hard part $T(x,Q)$ can be easily evaluated, and
the $\hat{\phi}$ defined at the scale $\mu\sim m$ is simply
$\delta(x-{1\over 2})$, to be compatible with the NRQCD matching
condition. Using the standard evolution equation to evolve this LCDA
into higher scale $\mu\sim Q$, combined with the tree-level result
of $T$, Eq.~(\ref{lc-factorization}) then automatically sums the
leading logarithms to all orders in $\alpha_s$.

In the remainder of the paper, we will test our understanding
through two explicit examples of exclusive production of a
bottomonium plus a photon. We will illustrate how the leading
kinematic logarithms in the NRQCD short-distance coefficient can be
identified and resummed within the framework of the factorization
formula (\ref{lc-factorization}).

{\it Example 1---\:$e^+e^-\to \eta_b \gamma$.} Let us first consider
the exclusive $\eta_b+\gamma$ production at high-energy
electron-positron collision, say, at LEP experiment. Similar
processes at $B$ factory environment have been studied in a
light-front model~\cite{Feldmann:1997te} and the NRQCD
approach~\cite{Chung:2008km}. To our purpose, it suffices to focus
on the decay of a highly virtual photon, $\gamma^*(Q)\to \eta_b(P)+
\gamma(k)$. According to the NRQCD factorization, one can express
the decay amplitude at the lowest order in $v$ as
\bqa
{\mathcal M}[\gamma^*\to \eta_b+\gamma] &=& {\widehat{\mathcal
M}}\big[\gamma^*\to b\bar{b}(^1S_0^{(1)},P)+\gamma \big]  {\langle
\eta_b| \psi^\dagger \chi|0 \rangle\over \sqrt{2N_c\,m}}\,,
\eqa
where $\psi^\dagger$ and $\chi$ are two-component Pauli spinor
fields. The factor associated with the NRQCD matrix element is
chosen for convenience, mainly to compensate the fact that the
particle states in the amplitude is relativistically normalized, but
the $\eta_b$ state in NRQCD matrix element conventionally assumes
non-relativistic normalization~\cite{Bodwin:1994jh}. It is worth
mentioning that, the factor $\langle \eta_b| \psi^\dagger \chi |0
\rangle/\sqrt{m}$ coincides with $f_{\eta_b}$, the $\eta_b$ decay
constant, to the lowest-order accuracy in $\alpha_s$ and $v$.

${\widehat{\mathcal M}}$, being the short-distance coefficient,
represents the quark amplitude when $\eta_b$ is replaced by the free
$b$ and ${\bar b}$ quarks that equally share the total momentum $P$,
and have the spin-color wave function
${|\uparrow\downarrow\rangle-|\downarrow \uparrow\rangle\over
\sqrt{2}}\times{\delta^{ij}\over \sqrt{N_c}}$, where $N_c=3$ is the
number of color and $i,\,j$ represent color indices. The calculation
is usually expedited by utilizing the covariant projection
method~\cite{Kuhn:1979bb}, though alternative methods also
exist~\cite{Braaten:1996jt,Bobeth:2007sh}. The result is
\bqa
{\widehat{\mathcal M}}[\gamma^*\to b\bar{b}(^1S_0^{(1)},P)+\gamma]
&=& \sqrt{2 N_c}\,{e^2 e_b^2 \over Q^2} \epsilon_{\mu\nu\alpha\beta}
\varepsilon^\mu_{\gamma^*} \varepsilon^{*\nu}_{\gamma} Q^\alpha
k^\beta\, F\left({m^2\over Q^2}\right)\,,
\eqa
where $e_b=-{1\over 3}$ is the fractional electric charge of $b$
quark. We introduce $F$ as a dimensionless form factor. It can be
simultaneously expanded in powers of $\alpha_s$ and $m^2/Q^2$:
\bqa
F\left({m^2\over Q^2}\right) &=& \sum_{n=0}^\infty \sum_{l=0}^{n}
C_{nl}\, \left({\alpha_s \over \pi} \right)^n \ln^l\left(Q^2\over
m^2\right)+ O\left({m^2\over Q^2}\right),
\label{Expansion:F:log}
\eqa
where $C_{nl}$ are constants. As explained before, the ``higher
twist" contributions suppressed by $1/Q^2$ are omitted. For this
simple process, the leading logarithm at the $n$th loop order is of
the form $\alpha_s^n \ln^n(Q^2/m^2)$, and our goal is to deduce
$C_{nn}$ analytically as well as resum their effects to all orders
in $\alpha_s$~\footnote{Note that the function $F$ is complex-valued
in this process. Its imaginary part can contain sub-leading
logarithms. However, insofar as the leading logarithms are
concerned, we are allowed to ignore the contributions from the
imaginary part.}.

The difficulty of NRQCD matching computation rapidly grows once
beyond the tree level, and the resulting analytic expressions of $F$
are generally rather cumbersome, often plagued with special
functions depending on $m^2/Q^2$ in a complicated way (for some
concrete examples on single and double charmonium production to
one-loop order, see Ref.~\cite{Hao:2006nf, Jia:2007hy,
Gong:2007db}). However, the expanded form as indicated in
(\ref{Expansion:F:log}) is always very simple. It is desirable if
$F$ can be reproduced in a more efficient way. As discussed earlier,
this is indeed possible, if we try to disentangle ``hard" and
``collinear" quanta in the NRQCD matching coefficients. For this
example, a factorization formula in complete analogy to the
$\pi\!-\!\!\gamma$ form factor is expected to hold:
\bqa
F\left({m^2\over Q^2}\right) &=& \int^1_0 dx \, T(x,Q,\mu)\,
\hat{\phi}(x,m,\mu) + O\left({m^2\over Q^2}\right)\,,
\label{F:collinear-factor}
\eqa
where $\mu$ is an arbitrary factorization scale that separates the
hard and collinear contributions. The hard function $T$ is identical
to what appears in the $\pi\!-\!\!\gamma$ form factor, which can be
expanded as $T= T^{(0)}+{\alpha_s\over \pi}T^{(1)}+\cdots$. The
tree-level result, $T^{(0)}(x,Q) = {1\over x}+{1\over 1-x}$, is well
known, and $T^{(1)}$ has also been available for a long
time~\cite{delAguila:1981nk,Braaten:1982yp}. The ``long-distance"
contribution is characterized by $\hat{\phi}$, the twist-2 LCDA of a
free $b\bar{b}$ pair in the spin-color state $^1S_0^{(1)}$ and with
total momentum $P$ halved by its two constituents. This LCDA admits
a definition in term of operator matrix
element~\cite{Chernyak:1983ej}:
\bqa
\hat{\phi}(x, m,\mu) &=& -{1\over \sqrt{2N_c}}\int \! {d w^-\over
2\pi} \, e^{-i x P^+ w^-} \,
 \langle b\bar{b}(^1S_0^{(1)},P)|\,\bar{b}(0,w^-,\mathbf{0}_\perp)
 \gamma^+ \gamma_5
 b(0)\,|0\rangle\,,
\eqa
where for simplicity we have suppressed the light-like gauge link.
This definition is boost invariant along the $\hat{z}$ axis. The
$\hat{\phi}$ can be systematically expanded perturbatively,
$\hat{\phi} =\hat{\phi}^{(0)}(x,m,\mu)+{\alpha_s\over
\pi}\hat{\phi}^{(1)}(x,m,\mu)+\cdots$, here $\mu$ can be identified
with the renormalization scale of the operator matrix element. Its
tree-level result is very simple, $\hat{\phi}^{(0)}(x,\mu\sim m)
\equiv \hat{\phi}^{(0)}(x)= \delta(x-{1\over 2})$, embodying the
NRQCD matching condition. For the investigations on
$\hat{\phi}^{(1)}$, one may consult
Ref.~\cite{Bell:2008er,Ma:2006hc}. Substituting the tree-level
expressions of $T^{(0)}$ and $\hat{\phi}^{(0)}$ into
(\ref{F:collinear-factor}), one then obtains $C_{00}=4$, which
agrees with the result derived from the standard NRQCD
matching~\cite{Chung:2008km}.

The invariance of physical amplitude about the choice of the
factorization scale $\mu$ leads to the RG equation, which is
conventionally referred to as Brodsky-Lepage (BL)
equation~\cite{Lepage:1979zb}. This equation governs the evolution
of the LCDA from one scale to another one, through which the
collinear logarithms can be resummed. The BL equation reads:
\bqa
 \mu^2 {\partial\over \partial \mu^2} \hat{\phi}(x,\mu)
 &=&
{\alpha_s(\mu^2)\over \pi} \, \int^1_0 \! dy \, {C_F\over
2}V_0(x,y)\,\hat{\phi}(y,\mu),
\label{BL-evolution:eqn:spin:zero}%
\eqa
where
\bqa
V_0(x,y) &=& \left[{1-x\over 1-y}\left(1+{1\over
x-y}\right)\theta(x-y)+ {x\over y}\left(1+{1\over
y-x}\right)\theta(y-x)\right]_+ \, ,
\label{BL-kernel-pseudoscalar}%
\eqa
is the kernel governing the evolution of LCDA of a helicity-zero
meson. $C_F={N_c^2-1\over 2 N_c}$ is the Casmir for $SU(N_c)$
fundamental representation, the subscript ``+" implies the familiar
``plus" prescription. This kernel can be interpreted as the
amplitude for a quark with fractional plus-momentum $x$ and
antiquark with $1-x$ to become a quark with fractional plus-momentum
$y$ and antiquark with $1-y$ by the exchange of one collinear gluon.

As is well known, the kernel $V_0$ admits the eigenfunctions $G_n(x)
\equiv x(1-x) C_n^{(3/2)}(2x-1)$, which are Gegenbauer polynomials
of order ${3\over 2}$ multiplied by the weight function $x(1-x)$:
\begin{subequations}
\bqa
\int^1_0 \!dy \, V_{0}(x,y) \, G_n(y) &=& -\gamma_n \, G_n(x),
\label{V0:eigenfunction}
\\
\gamma_n &=& {1\over 2} + 2 \sum_{j=2}^{n+1} {1\over j}- {1
 \over (n+1)(n+2)}\,.
\label{Gegenbauer:eigenvalue}
\eqa
\end{subequations}
It is a standard procedure to decompose the $\hat{\phi}(x)$ in the
basis of $G_n$:
\begin{subequations}
\bqa
\hat{\phi}(x,\mu) &=& \sum_{n=0}^\infty \hat{\phi}_n(\mu) \, G_n(x),
\\
\hat{\phi}_n(\mu) &=&  {4(2n+3)\over (n+1)(n+2)}\int_0^1 \! dx \,
C_n^{(3/2)}(2x-1)\,\hat{\phi}(x,\mu).
\label{}%
\eqa
\end{subequations}
At leading logarithmic accuracy, one can express the form factor as
the infinite sum of Gegenbauer moments:
\bqa
F\left({m^2\over Q^2}\right)_{\rm LL} &=& \int^1_0 \!dx \,
T^{(0)}(x)\, \hat{\phi}^{(0)}(x,Q^2)
= \sum_{n=0}^{\infty} \hat{\phi}_{2n}^{(0)}(Q^2).
\label{Gegenbauer:sum:etab}%
\eqa
All the odd Gegenbauer moments vanish. In deriving
(\ref{Gegenbauer:sum:etab}), we have made use of the fact
$\int^1_0\! dx\,C_{2n}^{(3/2)}(2x-1)=1$ for any integer $n$. Using
$C_{2n}^{(3/2)}(0)={(-1)^n(2n+1)!!\over (2n)!!}$, and evolving the
the moments from the low scale $\mu\sim m$ to $Q$, we obtain
\begin{subequations}
\bqa
 F\left({m^2\over Q^2}\right)_{\rm LL} &=& \sum_{n=0}^{\infty}
\hat{\phi}_{2n}^{(0)}\left({\alpha_s(Q^2)\over \alpha_s(m^2)
}\right)^{d_{2n}}\,,
\label{example:1:LLS:all:orders}
\\
\hat{\phi}_{2n}^{(0)} &=& 4(-1)^{n}(4n+3) {(2n-1)!!\over(2n+2)!!}\,.
\eqa
\end{subequations}
The anomalous dimension for each moment is $d_n=2\,C_F
\gamma_n/\beta_0$, where $\gamma_n$ is given in
(\ref{Gegenbauer:eigenvalue}). $\beta_0={11\over 3}N_c-{2\over
3}n_f$ is the one-loop coefficient of QCD $\beta$ function and $n_f$
is number of active quark flavors. The running strong coupling
constant is given by $\alpha_s(Q^2)= {4\pi \over \beta_0
\ln(Q^2/\Lambda^2_{QCD})}$. $\hat{\phi}_{2n}^{(0)}$ is an
alternating series and its magnitude declines very slowly ($\propto
1/\sqrt{n}$ asymptotically~\cite{Bell:2008er}), and the convergence
of the sum becomes extremely slow. Practically, one has to include
an extraordinary number of terms in the sum to warrant the
convergence of the result.

\begin{figure}[htb]
\centerline{
\includegraphics[height=5.1cm]{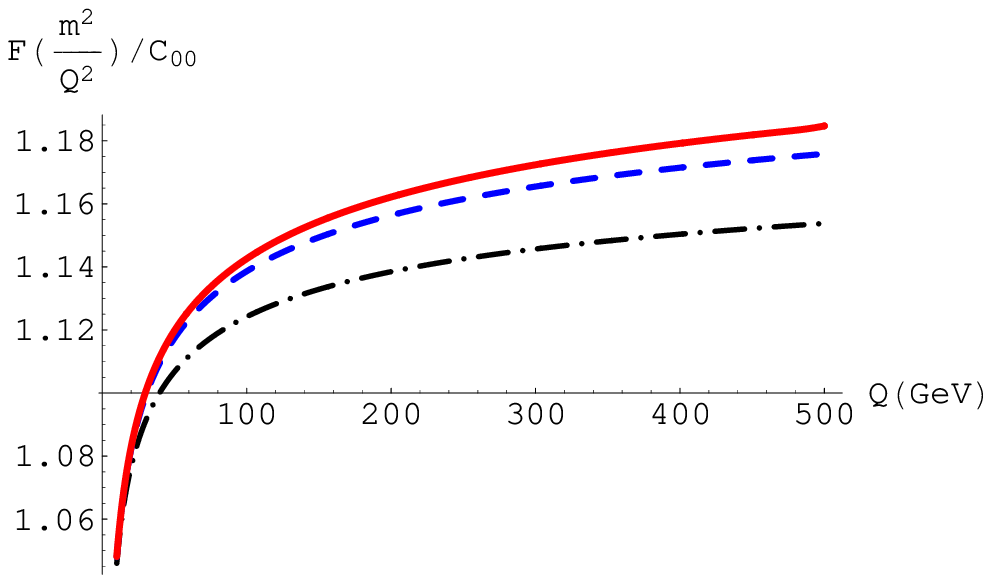}
\quad\quad
\includegraphics[height=5.1cm]{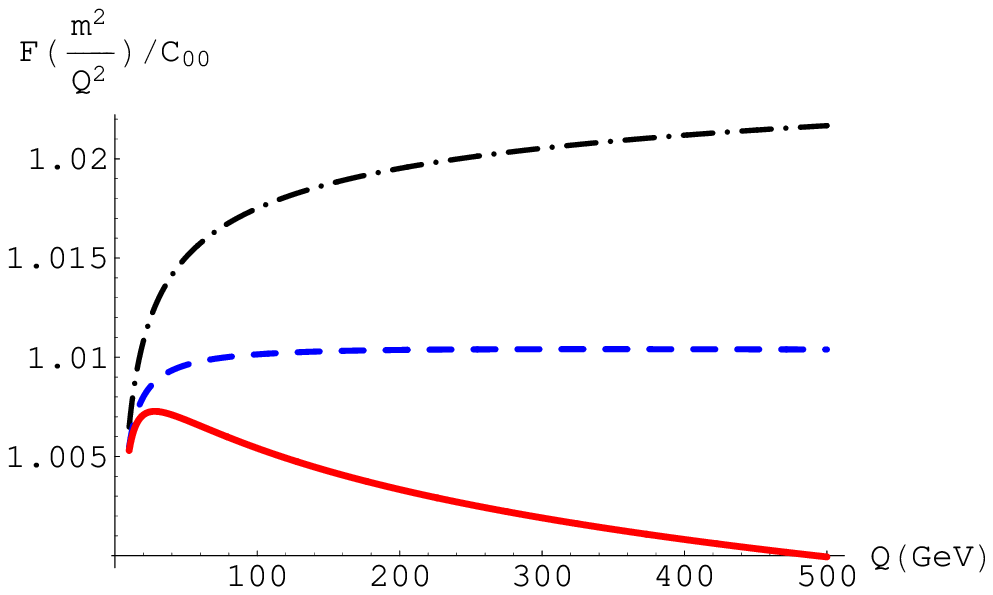}}
\caption{The form factor $F(m^2/Q^2)$ as a function of $Q$ for fixed
$m_b=4.7$ GeV. The left panel is for $\gamma^*\to \eta_b+\gamma$,
and the right panel for $h\to \Upsilon+\gamma$. We have fixed
$n_f$=5, $\beta_0={23\over 3}$ and $\Lambda_{\rm QCD}=100$ MeV.
Various curves represent the improved results to different extent in
leading logarithm approximation. The dot-dashed line only includes
the logarithm to one-loop order, the dashed line includes the
leading logarithms through two-loop order, and the solid line sums
the leading logarithms to all orders in $\alpha_s$.}
\label{fig:form:factor}
\end{figure}

In Fig.~\ref{fig:form:factor} we show the effect of leading
logarithm resummation, (\ref{example:1:LLS:all:orders}), for a wide
range of values of $Q$. One can clearly see that this RG-improved
result seems to only have a minor numerical impact, perhaps less
important than the constant terms in the fixed order calculation. As
$Q^2\to \infty$, only the first moment in
(\ref{example:1:LLS:all:orders}) can survive since $d_0=0$ and all
the other $d_{2n}$ are positive, therefore $F\to {3\over 2} C_{00}$.
However, as might be inferred from Fig.~\ref{fig:form:factor}, to
reach this asymptotic result, an unphysically large $Q$ ({\it i.e.}
$\gg M_{\rm Planck}$) is required.

Even though leading logarithms have been summed to all orders in
$\alpha_s$, it would still be illuminating if the coefficient of
each leading logarithm at $n$-th loop order, $C_{nn}$, can be
deduced. To this end, it is more convenient to work in the $x$ space
than the moment space. We can rewrite $F$ as:
\bqa
F\left({m^2\over Q^2}\right)_{\rm LL} &=& \int^1_0 \! dx\,
T^{(0)}(x) \hat{\phi}^{(0)}(x,Q)
=  \int^1_0 \! dx\, T^{(0)}(x)\, \exp[\kappa \,C_F \, V_0\,\star\,]
\,\hat{\phi}^{(0)}(x)\,,
\label{BL:schematic:x:space}
\eqa
where
\bqa
\kappa & \equiv & {2\over \beta_0}\ln\left({\alpha_s(m^2)\over
\alpha_s(Q^2) }\right)
=
{\alpha_s(Q^2)\over 2\pi} \ln \left({Q^2\over m^2}\right)+ \beta_0
{\alpha_s^2(Q^2)\over (4\pi)^2} \ln^2 \left({Q^2\over m^2}\right) +
\cdots\,.
\label{definition:kappa}%
\eqa
In the second equation of Eq.~(\ref{BL:schematic:x:space}), we have
substituted the symbolic solution of the BL equation for
$\hat{\phi}(x,Q)$, which is evolved from the infrared scale at
$\mu\sim m$ to the high scale $\mu\sim Q$.
The $\star$ operation implies that, upon the expansion of the
exponential, the resulting products of the argument of the
exponential are to be treated as convolutions. It is worth noting
that, Eq.~(\ref{BL:schematic:x:space}) is essentially the statement
that the moment-space amplitude exponentiates, as indicated in
(\ref{example:1:LLS:all:orders}).

We can expand equation (\ref{BL:schematic:x:space}) iteratively:
\bqa
F\left({m^2\over Q^2}\right)_{\rm LL} &=&  \int^1_0 \! dx \,
T^{(0)}(x)\,\hat{\phi}^{(0)}(x) + \kappa \, C_F \int^1_0 \! dx \!\!
\int^1_0 \! dy \, T^{(0)}(x)\, V_0(x,y) \hat{\phi}^{(0)}(y)
\nn \\
&+& {\kappa^2\,C_F^2\over 2!} \int^1_0 \! dx \!\! \int^1_0 \! dy
\!\! \int^1_0 \! dz \, T^{(0)}(x)\, V_0(x,y)\,V_0(y,z)\,
\hat{\phi}^{(0)}(z)+\cdots.
\label{iterative:solve:BL}%
\eqa
Equation (\ref{iterative:solve:BL}) admits a clear physical picture
how the leading logarithms are built up. At the hard vertex, $b$ and
$\bar{b}$ are produced to carry arbitrary fractional plus-momentum
$x$ and $1-x$, and transverse momenta of order $Q$. In the course of
evolution, they keep exchanging collinear gluons~\footnote{We have
tacitly made an important simplification. The kernel $V_0$ we use is
valid only for the flavor nonsinglet where annihilation of
$q\bar{q}$ is not allowed. In our case, it is possible that the
quark pair in the intermediate loop to annihilate into two gluons,
then recreates $b\bar{b}$ pair. These types of singlet diagrams
first occur at two loop. In this work, we neglect the logarithms
arising from such singlet contributions.} and reshuffle the
respective fractional plus-momenta, and gradually decrease the
transverse momenta. When ready to transition to a physical $\eta_b$
state, they finally adjust themselves to become nearly collinear and
halve the total momentum.

We choose the order of the multiple integration in
(\ref{iterative:solve:BL}) from the left to the right, and leaves
the integration over the $\hat{\phi}^{(0)}$ in the last step. With
the help of (\ref{T0:V0}) and (\ref{T0:V0:V0}), we are able to
deduce the leading logarithms through the two-loop order:
\bqa
F\left({m^2\over Q^2}\right)_{\rm LL} &=& C_{00}\left\{ 1+{C_F
\alpha_s(Q^2)\over 4\pi} \ln \left({Q^2\over m^2}\right)(3- 2 \ln 2)
\right.
\label{example:1:LL:2:loop}
\\
&+&  \left. C_F {\alpha_s^2(Q^2)\over (4\pi)^2} \ln^2
\left({Q^2\over m^2}\right) \left[\beta_0 \left({3\over 2}-\ln 2
\right) + C_F\left({9\over 2}-{\pi^2\over 6}-8 \ln2+ \ln^2 2
\right)\right]+ \cdots\right\}\,,
\nn
\label{two:loop:LL}%
\eqa
where the coefficient $C_{11}$ agrees with the asymptotic expansion
of the exact NLO correction to the NRQCD short-distance
coefficient~\cite{Sang:2008}, and can also be read off by imposing
$x={1\over 2}$ in $T^{(1)}(x)$ in Ref.~\cite{Braaten:1982yp}. Our
prediction to $C_{22}$ is new. From Fig.~\ref{fig:form:factor}, one
can see that, for a wide range of $Q/m$, the leading logarithm
approximation through two-loop order is already quite close to the
resummed results, (\ref{example:1:LLS:all:orders}).

{\it Example 2---Higgs radiative decay into $\Upsilon$}. It is
tempting to test our understanding in a slightly different
situation, {\it e.g.} a process that a transversely polarized
$\Upsilon$ is produced at leading power. Let us consider $h(Q)\to
\Upsilon(P)+ \gamma(k)$. In view of its clean signature, this decay
channel may be potentially useful to search for the Higgs boson in
the forthcoming CERN Large Hadron Collider (LHC) experiment, were it
not completely swallowed by the much more copious background. In
accordance with the NRQCD factorization, we can express the
amplitude at the lowest order in $v$ as~\footnote{Our main concern
here is for the illustrative purpose, so we do not target at a
serious phenomenological analysis of this decay channel. For
instance, we have not included the fragmentation contribution from
$h\to \gamma^*\gamma\to \Upsilon\gamma$.}
\bqa
{\mathcal M}[h\to \Upsilon+\gamma]  &=& {\widehat{\mathcal M}}\big[h
\to b\bar{b}(^3S_1^{(1)},P)+\gamma \big]  {\langle
\Upsilon(\bm{\epsilon})| \psi^\dagger {\bm\sigma} \cdot
\bm{\epsilon}\chi|0 \rangle\over \sqrt{2N_c\,m}}.
\eqa
It might be worth pointing out that, the NRQCD matrix element
$\langle \Upsilon(\bm{\epsilon})| \psi^\dagger {\bm\sigma} \cdot
\bm{\epsilon}\chi|0 \rangle/\sqrt{m}$ coincides with $f_{\Upsilon}$,
$\Upsilon$ decay constant, to the lowest-order accuracy in
$\alpha_s$ and $v$.

Analogous to the first example, ${\widehat{\mathcal M}}$ here
represents the parton amplitude when $\Upsilon$ is replaced by the
free $b$ and ${\bar b}$ quarks in the $^3S_1^{(1)}$ state, each of
which carries half of the total momentum $P$. It can be expressed as
follows:
\bqa
{\widehat{\mathcal M}}\big[h \to b\bar{b}(^3S_1^{(1)},P)+\gamma
\big] &=& -\sqrt{2 N_c}\,{e\,e_b\over 2} \, g_{h b\bar{b}}\,
\varepsilon^*_{\Upsilon}\cdot \varepsilon^*_{\gamma} \,
F\left({m^2\over Q^2}\right),
\eqa
where $Q^2=M_h^2$, $g_{hb\bar{b}}=m/{\textsf v}$ is the Yukawa
coupling between Higgs and $b$ quarks, and $\textsf{v}=246 $ GeV
implies the vacuum expectation value of the Higgs field. Since the
photon must be transversely polarized, so is the $\Upsilon$.

The dimensionless form factor $F$ here is expected to also satisfy a
collinear factorization formula similar to
(\ref{F:collinear-factor}). The tree-level hard coefficient function
$T^{(0)}$ is the same as the preceding example, but here one should
use the twist-2 LCDA of the transversely polarized
$b\bar{b}(^3S_1^{(1)},P)$ state, which can be defined as the
following operator matrix element~\cite{Chernyak:1983ej}:
\bqa
\hat{\phi}_\perp(x, m,\mu) &=&  {i\over \sqrt{2N_c}}\int \! {d
w^-\over 2\pi} \, e^{-i x P^+ w^-}
\nn \\
&\times & \langle b\bar{b}(^3S_1^{(1)}, P, \lambda_\perp)|\,
 \bar{b}(0,w^-,\mathbf{0}_\perp)\,
\sigma^{+\mu} \epsilon_\mu(\lambda_\perp)\,
 b(0)\,|0\rangle\,.
\eqa
At the lowest order in $\alpha_s$, $\hat{\phi}_\perp^{(0)}(x,\mu\sim
m)\equiv \hat{\phi}^{(0)}(x)=\delta(x-1/2)$. Thus in this example,
we also get $C_{00}=4$. The $\hat{\phi}_\perp$ also satisfies the BL
equation (\ref{BL-evolution:eqn:spin:zero}), except the kernel $V_0$
should be replaced by a different one:
\bqa
V_\perp(x,y)  &=& V_0(x,y)-\left[{1-x\over 1-y}\, \theta(x-y)+
{x\over y}\, \theta(y-x) \right]\,.
\label{BL-kernel-Vector-perp}%
\eqa
This new kernel permits the same eigenfunctions $G_n(x)$ as given in
Eq.~(\ref{V0:eigenfunction}), but the corresponding eigenvalues are
slightly different from Eq.~(\ref{Gegenbauer:eigenvalue}), {\it
i.e.} $\widetilde{\gamma}_n ={1\over 2} + 2 \sum_{j=2}^{n+1} {1\over
j}$.

Completely analogous to the preceding example, we can express the
form factor $F$ as the infinite sum of Gegenbauer moments, with
evolution effect taken into account:
\bqa
F\left({m^2\over Q^2}\right)_{\rm LL} &=& \int^1_0 \!dx
\,T^{(0)}(x)\, \hat{\phi}_\perp^{(0)}(x,Q^2)
\nn\\
&=& \sum_{n=0}^{\infty} \hat{\phi}_{\perp\:2n}(Q^2) =
\sum_{n=0}^{\infty} \hat{\phi}_{2n}^{(0)}\left({\alpha_s(Q^2)\over
\alpha_s(m^2) }\right)^{{\tilde d}_{2n}},
\label{example:2:LLS:all:orders}%
\eqa
where the anomalous dimension for the moment is
$\tilde{d}_n=2\,C_F\,\widetilde{\gamma}_n/\beta_0$.

We also show the effect of leading logarithm resummation,
(\ref{example:2:LLS:all:orders}), for a wide range of values of $Q$
in Fig.~\ref{fig:form:factor}. One can clearly see that this
RG-improved result seems to have an even minor impact compared with
the preceding example~\footnote{Our investigations on these two
examples show that, even when $Q\gg m$, leading logarithm
resummation may not play a significant role. This finding provides a
counterexample to the assertion made in \cite{Braguta:2008tg}.}.
Since all the $\tilde{d}_{2n}$ are positive, as $Q^2\to \infty$, one
expects $F\to 0$. However, as indicated in
Fig.~\ref{fig:form:factor}, to reach this asymptotic result, an
unphysically large Higgs mass is required.

We can also iteratively derive the closed form for the leading
logarithms occurring at the $n$-th loop order, by simply replacing
the kernel $V_0$ in (\ref{iterative:solve:BL}) by $V_\perp$. Using
(\ref{T0:Vperp}) and (\ref{T0:Vperp:Vperp}), we can predict the
leading logarithms through the two-loop order:
\bqa
F\left({m^2\over Q^2}\right)_{\rm LL} &=& C_{00}\left\{ 1+{C_F
\alpha_s(Q^2)\over 4\pi} \ln \left({Q^2\over m^2}\right)(3- 4\ln 2)
\right.
\label{example:2:LL:2:loop}
\\
&+&  \left. C_F {\alpha_s^2(Q^2)\over (4\pi)^2} \ln^2
\left({Q^2\over m^2}\right) \left[\beta_0 \left( {3\over 2}-2\ln 2
\right) + C_F \left({9 \over 2}-12 \ln 2+ 4\ln^2 2 \right)\right]+
\cdots\right\}.
\nn
\label{example2:two:loop:LL}%
\eqa
As a check, we have explicitly computed the logarithmical
contribution in $T^{(1)}({1\over 2})$ for this
process~\footnote{There is a subtlety arising in this process. Since
the composite operator $\bar{b} b$ acquires an anomalous dimension,
the renormalization procedure has to be performed to the $T^{(1)}$,
where a renormalization scale $\mu_R$ will enter and give rise to
the logarithm like $\ln(Q^2/\mu_R^2)$. This type of logarithm has a
very different origin from the collinear logarithm we are
interested, so will be eliminated by choosing $\mu_R\sim Q$.}. It
agrees with $C_{11}$ given above. As one can tell from
Fig.~\ref{fig:form:factor}, even though including the leading
logarithm at two-loop order noticeably modifies the one-loop result,
the overall effect of logarithms is too modest to be
phenomenologically relevant.

\emph{Summary.} In this paper we suggest that, for a class of
exclusive single-quarkonium production processes, the short-distance
coefficient in the leading order NRQCD expansion can be further
separated into ``hard" and ``collinear" degrees of freedom. As a
consequence, the corresponding NRQCD matching coefficient can be
expressed as the convolution between a hard coefficient function and
a perturbatively calculable LCDA of a properly-chosen free
quark-antiquark pair. The procedure of refactorization has both
conceptual and technical advantages over the conventional NRQCD
matching calculation. For instance, we have shown that large
kinematic logarithms can be readily summed to all orders in
$\alpha_s$ within this scheme, which is otherwise difficult to
accomplish. This strategy also provides an efficient means to
reproduce the NRQCD matching coefficients when proceeding beyond the
tree level (for an explicit one-loop illustration, see
~\cite{Jia:2009}).

We believe that the idea of refactorization is rather general and
deserves to be further explored in other more complicated cases. For
example, the NRQCD short-distance coefficients associated with
exclusive $P$-wave quarkonium production and relativistic
corrections to $S$-wave quarkonium production, should also be
amenable to a similar collinear factorization formula. It is
conceivable that exclusive double charmonium production processes
like $e^+ e^-\to J/\psi+\eta_c$, may even be approachable from this
angle. However, since such processes often violate the helicity
selection rule and the amplitude usually starts at subleading power
of $m^2/Q^2$, consequently higher-twist LCDAs must be
introduced~\cite{Ma:2004qf,Bondar:2004sv,Braguta:2008tg}. This may
pose some great challenge to fulfill factorization and evolution.

\acknowledgments

We thank Wen-Long Sang for providing us with the asymptotic
expression of the NLO QCD correction to the process $e^+ e^- \to
\eta_c\gamma$. This work was supported in part by the National
Natural Science Foundation of China under Grants No.~10605031,
10705050 and 10875130.

{\it Note added.} After the manuscript was submitted, we became
aware of a work done by Shifman and Vysotsky in
1981~\cite{Shifman:1980dk}, who had considered two basically
identical processes as ours, and also attempted to resum the
respective leading kinematic logarithms. Though differing in
technical taste, their and our approaches are essentially the same,
both derived from the light-cone OPE. Nevertheless, our approach
seems easier to follow in practice.
For the first process $e^+e^-\to \eta_b\gamma$, they also considered
the mixing effects due to the two-gluon Fock components of $\eta_b$,
which we have not. Both of works agree on the one-loop coefficient
of leading logarithms in this process. Our approach enables one to
readily deduce the two-loop coefficient as well, but it would be a
laborious undertaking for their formalism to achieve this.
For the second process $h\to \Upsilon\gamma$, they combined the
collinear logarithms with those logarithms arising from the running
quark mass. After subtracting the latter type of logarithms, their
results agree with ours.
It should be noted, however, the strength of refactorization
proposed in this work is not only limited to summing large
logarithms. As has been stressed throughout the paper, our approach
will serve an efficient and systematic tool to reproduce the NRQCD
short-distance coefficients to any desired order in $\alpha_s$,
including the logarithms as well as the constants, in various
exclusive quarkonium production processes. In this regard, we feel
that our work has gone one step farther than
Ref.~\cite{Shifman:1980dk}.

\appendix

\section{Useful mathematical formulas
\label{appendix} }

In {\it Example 1}, in order to iteratively deduce the leading
logarithms through the two-loop order from
(\ref{iterative:solve:BL}), we need know the following integrals:
\begin{subequations}
\bqa
&& \int^1_0 \! dx \, T^{(0)}(x)\, V_0(x,y)= {1\over
y(1-y)}\left[{3\over 2}+(1-y)\ln y+ y \ln(1- y)\right],
\label{T0:V0}
\\
&&\int^1_0 \! dx \!\! \int^1_0 \! dy\, T^{(0)}(x)\, V_0 (x,y)\, V_0
(y,z) = {1\over z(1-z)} \left[ {9\over 4}-{\pi^2\over 6}+\ln z +
\ln(1-z) +  2(1-z)\ln z \right.
\nn\\
&& \left.+ 2z \ln(1-z)+(1-z)\ln^2 z + z\ln^2(1-z) + z\,{\rm
Li}_2(z)+ (1-z){\rm Li}_2(1-z) \right],
\label{T0:V0:V0}%
\eqa
\end{subequations}
where ${\rm Li}_2$ denotes dilogarithm. Eq.~(\ref{T0:V0}) can be
found in Ref.~\cite{Braaten:1982yp}, and (\ref{T0:V0:V0}) is new.

Expand the Gegenbauer-moment-summation formula
(\ref{example:1:LLS:all:orders}) in $d_{2n}$, truncate the power
series in $\alpha_s(Q^2)\ln (Q^2/m^2)$ to the second order, and
match them onto Eq.~(\ref{example:1:LL:2:loop}). Enforcing the
mutual equivalence leads to the following mathematical identities:
\begin{subequations}
\bqa
\sum_{n=0}^{\infty} \hat{\phi}_{2n}^{(0)} &=& 4.
\\
\sum_{n=0}^{\infty} \gamma_{2n} \,\hat{\phi}_{2n}^{(0)} &=& -2
(3-2\ln 2).
\\
\sum_{n=0}^{\infty} \gamma^2_{2n} \,\hat{\phi}_{2n}^{(0)} &=&
9-{\pi^2\over 3}-16 \ln2+ 2 \ln^2 2 \,.
\eqa
\end{subequations}

Analogously, in {\it Example 2}, to iteratively acquire the leading
logarithms up to two-loop order, we need to know the following
integrals:
\begin{subequations}
\bqa
&& \int^1_0 \! dx \,T^{(0)}(x)\, V_\perp(x,y)= {1\over
y(1-y)}\left[{3\over 2}+\ln y + \ln(1- y)\right],
\label{T0:Vperp}
\\
&&\int^1_0 \! dx \!\! \int^1_0 \!\! dy\, T^{(0)}(x)\, V_\perp(x,y)
V_\perp(y,z)
\nn\\
&=&  {1\over z(1-z)} \left[ {9\over 4}+3\ln z + 3\ln(1-z) + \ln^2 z
+ \ln^2(1-z) \right]\,.
\label{T0:Vperp:Vperp}
\eqa
\end{subequations}

Expand the Gegenbauer-moment-summation formula
(\ref{example:2:LLS:all:orders}) in $\tilde{d}_{2n}$, truncate the
power series in $\alpha_s(Q^2)\ln (Q^2/m^2)$ to the second order,
and compare them with (\ref{example:2:LL:2:loop}). The mutual
equivalence demands that the following identities must hold:
\begin{subequations}
\bqa
\sum_{n=0}^{\infty} \widetilde{\gamma}_{2n}\, \hat{\phi}_{2n}^{(0)}
&=& -2 (3-4\ln 2)\,,
\\
\sum_{n=0}^{\infty}  \widetilde{\gamma}_{2n}^2
\,\hat{\phi}_{2n}^{(0)} &=& 9 - 24 \ln2+ 8 \ln^2 2\,.
\eqa
\end{subequations}


\end{document}